\documentclass[aps, prd, twocolumn, lengthcheck, superscriptaddress, 
nofootinbib]{revtex4-1}

\usepackage{amsmath,amssymb,amsfonts,physics,float}
\usepackage{booktabs} 
\usepackage[setpagesize=false]{hyperref}
\allowdisplaybreaks[4]
\usepackage{graphicx}
\usepackage[labelformat=simple]{subcaption} 

\usepackage{ulem}
\usepackage{color}
\definecolor{ar}{rgb}{1.0, 0.01, 0.24}
\definecolor{al}{rgb}{0.82, 0.1, 0.26}
\definecolor{ev}{rgb}{0.56, 0.0, 1.0}

\def\be{\begin{eqnarray}}
\def\ee{\end{eqnarray}}

\begin{document}

\title{
Origin of nucleon mass in the light of PSR J0614-3329 with quark-hadron crossover
}

\author{Bikai Gao}
\email{bikai@rcnp.osaka-u.ac.jp}
\affiliation{Research Center for Nuclear Physics (RCNP), Osaka University, Osaka 567-0047, Japan}

\author{Yuk-Kei Kong}
\email{yukkekong2-c@hken.phys.nagoya-u.ac.jp}
\affiliation{Department of Physics, Nagoya University, Nagoya 464-8602, Japan}

\author{Yong-Liang Ma}
\email{ylma@nju.edu.cn}
\affiliation{School of Frontier Sciences, Nanjing University, Suzhou, 215163, China}

\date{\today}

\begin{abstract}
The recent NICER observation of PSR J0614-3329, revealing the smallest reliably measured neutron star radius of $R = 10.29^{+1.01}_{-0.86}$ km at mass $M = 1.44^{+0.06}_{-0.07} M_\odot$, provides an unprecedented constraint on the equation of state of dense matter. We investigate the implications of this measurement for the origin of nucleon mass within the parity doublet model framework, which naturally incorporates both chiral variant and chiral invariant mass components. We construct unified equations of state by employing the parity doublet model with isovector scalar meson $a_0(980)$ for hadronic matter up to twice nuclear saturation density, smoothly connected to a Nambu-Jona-Lasinio-type quark model at higher densities through a crossover transition. By systematically varying the chiral invariant mass $m_0$ and quark matter parameters, we determine which values simultaneously satisfy all current astrophysical constraints, including gravitational wave observations from GW170817, NICER measurements of several pulsars, and the existence of two-solar-mass neutron stars. The inclusion of PSR J0614-3329 dramatically refines the allowed range of the chiral invariant mass from the previous constraint of $580~\text{MeV} \lesssim m_0 \lesssim 860~\text{MeV}$ to $800~\text{MeV} \lesssim m_0 \lesssim 860~\text{MeV}$, raising the lower bound by approximately 220 MeV. This result indicates that the chiral invariant mass must constitute at least 85$\%$ of the nucleon mass, challenging the traditional picture of nucleon mass generation through spontaneous chiral symmetry breaking alone and highlighting the importance of gluon condensation and other non-chiral mechanisms. 
\end{abstract}

\maketitle


\section{Introduction}
Neutron stars (NSs) provide natural laboratories for investigating cold, dense matter under extreme conditions inaccessible to terrestrial experiments. These compact objects, with typical masses of $1$--$2 M_\odot$ and radii of $10$--$12$ km, harbor matter at densities exceeding several times nuclear saturation density ($n_0 \simeq 0.16$ fm$^{-3}$). The macroscopic properties of NSs—their masses, radii, tidal deformabilities, and moment of inertia—are directly connected to the microscopic interactions governing their interior matter. This intimate connection offers crucial insights into many fundamental aspects of nuclear physics, including the Quantum Chromo Dynamics (QCD) phase diagram at finite density, the nature of quark-hadron phase transitions, the emergence of exotic degrees of freedom, and the behavior of fundamental symmetries under extreme conditions.

Recent breakthroughs in multi-messenger astronomy have dramatically improved our ability to constrain the NS equation of state (EOS). The detection of gravitational waves from the binary NS merger GW170817 by the LIGO-Virgo collaborations~\cite{PhysRevLett.119.161101,LIGOScientific:2017ync,LIGOScientific:2018cki} provided the first constraints on NS tidal deformability, favoring relatively soft EOSs at intermediate densities. Complementing these gravitational wave observations, NASA's Neutron Star Interior Composition Explorer (NICER) has achieved remarkable precision in measuring NS masses and radii through X-ray pulse profile modeling. Observations of PSR J0030+0451~\cite{Riley:2019yda, Vinciguerra:2023qxq} and PSR J0437-4715~\cite{Choudhury:2024xbk} have collectively constrained the radius of a $1.4 M_\odot$ NS to $R_{1.4} \approx 11$--$13$ km. Additionally, the detection of massive pulsars such as PSR J0740+6620 with masses near $2 M_\odot$~\cite{Riley:2021pdl,Miller:2021qha,Raaijmakers:2021uju,Dittmann:2024mbo,Salmi:2024aum} requires the EOS to be sufficiently stiff at high densities.

These observational constraints on the EOS are fundamentally connected to the behavior of nucleon mass at finite densities. Traditionally, the nucleon mass of approximately 940 MeV has been understood to emerge from the spontaneous breaking of chiral symmetry in QCD, which generates a non-zero quark condensate $\langle\bar{q}q\rangle \approx -(250~\text{MeV})^3$. In this conventional picture, as density increases, the quark condensate diminishes, leading to partial restoration of chiral symmetry and a corresponding reduction in nucleon mass. However, recent lattice QCD simulations at finite temperature~\cite{Aarts:2015mma,Aarts:2017rrl,Aarts:2017iai,Aarts:2019dot} and QCD sum rule calculations~\cite{Kim:2020zae,Kim:2021xyp,Lee:2023ofg} have revealed that the nucleon mass is not completely from the chiral symmetry breaking. Instead, the nucleon mass should posses two distinct components: a density-dependent "chiral variant" mass arising from quark condensation and spontaneous chiral symmetry breaking, and a chiral invariant mass $m_0$ that persists even in the chiral symmetry restored phase, generated by gluon condensation and/or multi-quark condensation. This chiral invariant mass represents a fundamental parameter that determines the nucleon mass and has profound implications for the EOS at high densities relevant to NS cores, for example, how the dilaton-limit fixed point is approached~\cite{Paeng:2011hy,Ma:2019ery}. The parity doublet model (PDM)~\cite{Detar:1988kn,Jido:2001nt} provides a natural framework for incorporating both mass components, offering a more complete description of how nucleon properties evolve with density---essential for constructing  EOSs and interpreting NS observations.

Previous studies have successfully applied the PDM to NS matter, demonstrating that the chiral invariant mass significantly affects the EOS stiffness and NS radii~\cite{Zschiesche:2006zj, Dexheimer:2007tn, Dexheimer:2008cv, Sasaki:2010bp, Sasaki:2011ff,Steinheimer:2011ea, Motohiro:2015taa, Mukherjee:2016nhb,  Gallas:2011qp, Suenaga:2017wbb,Marczenko:2017huu,  Marczenko:2018jui, Yamazaki:2019tuo, Marczenko:2019trv, Minamikawa:2021fln,Gao:2024mew, Gao:2024chh, Gao:2024jlp,Yuan:2025dft,Kong:2025dwl,Gao:2025nkg}. Furthermore, unified EOSs constructed by combining the PDM for hadronic matter with Nambu--Jona-Lasinio (NJL) type models~\cite{Hatsuda:1994pi, Buballa:2003qv, Fukushima:2008wg, Song:2019qoh,Baym:2017whm,Xia:2024juh,Gholami:2024diy,Kawaguchi:2024edu,Christian:2025dhe} for color flavor locked quark matter~\cite{Rajagopal:2000wf,Alford:2003fq,Alford:2007xm,Alford:2017qgh,Oikonomou:2023otn,Chu:2024kdf,Ivanytskyi:2024zip} through a quark-hadron crossover~\cite{Minamikawa:2021fln,Minamikawa:2023eky,Gao:2024mew} suggest that the chiral invariant mass should exceed half of the nucleon mass. A similar conclusion was obtained in the skyrmion crystal approach to nuclear matter~\cite{Ma:2013ooa,Ma:2013ela}.

Most recently, NICER observations of PSR J0614-3329 have provided a crucial new constraint that challenges our understanding of dense matter. This measurement yields the smallest reliably determined NS radius to date: $R = 10.29^{+1.01}_{-0.86}$ km at mass $M = 1.44^{+0.06}_{-0.07} M_\odot$~\cite{Mauviard:2025dmd}. This unexpectedly small radius implies substantial softening of the EOS at intermediate densities, while the existence of $2M_\odot$ NSs still requires sufficient stiffness at high densities. This tension makes it increasingly challenging to satisfy all observational constraints simultaneously within theoretical frameworks~\cite{Tang:2025xib,Shirke:2025gfi}.

In this work, we investigate the implications of PSR J0614-3329 for the chiral invariant mass $m_0$ within the PDM framework, aiming to establish quantitative connections between this astrophysical observation and the fundamental question of nucleon mass origin. We construct unified EOSs by employing the PDM with isoscalar scalar meson $\sigma$, isovector scalar $a_0$ meson, isoscalar vector meson $\omega$ and isovector vector meson $\rho$ contributions for densities up to $2n_0$, then smoothly interpolate to an NJL-type quark model at higher densities through a crossover transition. By systematically varying $m_0$ and other model parameters, we determine which values are consistent with the full set of NS observations, including the stringent new constraint from PSR J0614-3329, and found that $m_0$ is constrained to $800~\mbox{MeV} \lesssim m_0 \lesssim 860~\mbox{MeV}$, about $85\%$ of the nucleon mass. Our analysis provides insights into how recent astrophysical observations constrain fundamental properties of QCD matter.

The paper is organized as follows: In Sec.~\ref{sec-eos}, we present the theoretical framework for constructing the EOS. In Sec.~\ref{sec-NS}, we construct unified EOS across the whole density region and analyze NS properties. Finally, Sec.~\ref{summary} summarizes our main findings and discusses their implications for understanding the fundamental origin of nucleon mass in QCD.


\section{EQUATION OF STATE }
\label{sec-eos}

In this section, we briefly review the construction of neutron star matter EOS using PDM for the low-density hadronic region and a NJL-type quark model for the high-density region. 

\subsection{NUCLEAR MATTER EOS}
\label{sec:PDM matter}

For the low-density region ($n_B \lesssim 2n_0$), the hadronic model based on parity doublet structure was developed in Ref.~\cite{Gao:2022klm}. The model includes the effects of strange quark chiral condensate through the KMT-type interaction in the mesonic sector. 
This model was originally constructed to include strange quark effects through KMT-type interactions in the mesonic sector. However, detailed calculations of the density-dependent strange quark chiral condensate $\langle\bar{s}s\rangle$ revealed that its impact on NS properties remains negligible at low densities. Therefore, we neglect strange quark contributions in the present analysis of the hadronic phase.

To ensure the consistency with current nuclear physics constraints, we incorporate two important modifications to the standard PDM framework. First, we include a vector meson mixing term ($\omega^2\rho^2$) to reproduce the empirical slope parameter $L = 57.7 \pm 19$ MeV of the symmetry energy, as constrained by recent analyses~\cite{universe7060182}. Second, we incorporate the isovector scalar meson $a_0(980)$ to provide a more complete description of isospin-dependent interactions. The $a_0(980)$ meson plays a crucial role in nuclear matter by mediating attractive forces in the isovector channel.
Its effects on the symmetry energy and the EOS of asymmetric matter have been extensively studied within various theoretical frameworks. Investigations using Walecka-type relativistic mean-field (RMF) models~\cite{Kubis_1997,Zabari:2018tjk,Kubis:2020ysv,Miyatsu_2022,Li_2022,Thakur_2022,Liu_2005,Ma:2023eoz,Ma:2025llw,PhysRevC.80.025806,Gaitanos_2004,PhysRevC.67.015203,PhysRevC.65.045201} and density-dependent RMF approaches~\cite{PhysRevC.90.055801,PhysRevC.84.054309} consistently demonstrate that including the $a_0(980)$ meson increases the symmetry energy
and stiffens both the NS EOS
and the EOS of asymmetric matter.
These effects become particularly important for accurately describing NS radii and tidal deformabilities.

The thermodynamic potential of the nucleon part is given by 
\begin{equation}
\begin{aligned}
{} & \Omega_{N}  = - 2 \sum_{\alpha=\pm, j=\pm} \int^{k_f} \frac{d^3p}{(2 \pi)^3} \bigg[ \mu^*_j - \omega_{\alpha j} \bigg],
\end{aligned}
\label{OFG}
\end{equation} 
where $\alpha = \pm$ denotes the parity and $j = \pm$ denotes the isospin of nucleons ($j=+$ for proton and $j=-$ for neutron). The effective chemical potential $\mu^*_j$ incorporates the contributions from vector mesons
\begin{equation}
\begin{aligned}
\mu^*_j \equiv (\mu_B - g_{\omega NN} \omega) + \frac{j}{2} (\mu_I - g_{\rho NN} \rho)\ .
\end{aligned}
\label{mustar}
\end{equation} 
where $g_{\omega NN}$ and $g_{\rho NN}$ are the coupling constants to be determined. The single-particle energy $\omega_{\alpha j} = \sqrt{(\vec{p})^2 + (m^*_{\alpha j})^2}$ depends on the momentum $\vec{p}$ and the effective mass $m^*_{\alpha j}$, which in the PDM framework takes the form 
\begin{equation}
\begin{aligned}
  m^*_{\alpha j} = \frac{1}{2} \bigg[ \sqrt{(g_1+g_2)^2(\sigma - ja)^2 + 4m_0^2} \\
  + \alpha(g_1 - g_2)(\sigma - ja) \bigg]\ .
\end{aligned}
\label{maj}
\end{equation}
with $g_1, g_2$ the Yukawa coupling constants,  $m_0$ is the chiral invariant mass and $\sigma, a$ is obtain from the mean-field approximation 
\begin{align}
\sigma(x) \rightarrow \sigma, \quad a_0^i(x) \rightarrow a \delta_{i 3}.
\end{align}

The complete hadronic thermodynamic potential includes contributions from both nucleons and mesons
\begin{equation}
\begin{aligned}
 \Omega_{\rm PDM}  =&\Omega_{N}+  V(\sigma, a) - V_0(\sigma, a) \\
&   - \frac{1}{2} m^2_{\omega} \omega^2 - \frac{1}{2} m^2_{\rho} \rho^2 \\
&- \lambda_{\omega \rho} g_{\omega NN}^2 g_{\rho NN}^2 \omega^2 \rho^2,
\end{aligned}
\label{eq36m}
\end{equation} 
where the mesonic potential $V(\sigma, a)$ contains both quadratic and higher-order terms
\be
V(\sigma, a) & = &{} - \frac{\bar{\mu}^2_{\sigma}}{2} \sigma^2 - \frac{\bar{\mu}^2_{a}}{2} a^2 + \frac{\lambda_4}{4} (\sigma^4 + a^4 ) + \frac{\gamma_4}{2} \sigma^2 a^2 \nonumber\\
& &{} - \frac{\lambda_6}{6} (\sigma^6 +15\sigma^2a^4 + 15\sigma^4a^2+ a^6 ) \nonumber\\
& &{} + \lambda_6^{\prime}(\sigma^2a^4 + \sigma^4a^2) - m_\pi^2f_\pi \sigma,
\ee
and the vacuum potential
\begin{equation}
\begin{aligned}
V_0(\sigma, a) = - \frac{\bar{\mu}^2_{\sigma}}{2} f_{\pi}^2 + \frac{\lambda_4}{4} f_{\pi}^4 
 - \frac{\lambda_6}{6} f_{\pi}^6 - m^2_{\pi} f_{\pi}^2\ .
\end{aligned}
\label{eq37}
\end{equation}      
is subtracted to ensure proper normalization. Following the analysis in Refs.~\cite{PhysRevC.108.055206,sym160912382024}, terms with coefficient $\lambda^\prime_6$ are sub-leading in the large-$N_c$ expansion and have negligible effects on matter properties. We therefore set $\lambda^\prime_6 = 0$ for simplicity. Additionally, vacuum stability requires $\lambda_6 > 0$~\cite{Kong:2023nue}. For neutron star matter, we include leptonic contributions to obtain the total thermodynamic potential
\begin{align}
\Omega_{{\rm H}} = \Omega_{{\rm PDM}} + \sum_{l = e, \mu}\Omega_l \ , 
\end{align}
where $\Omega_{l}(l=e,\mu)$ are the thermodynamic potentials for leptons  given by 
\begin{equation}
\Omega_{l}=-2 \int^{k_{F}} \frac{d^{3} \mathbf{p}}{(2 \pi)^{3}}\left(\mu_{l}-E_{\mathbf{p}}^{l}\right).
\end{equation}
The mean fields here are determined by following stationary conditions:
\begin{equation}
\frac{\partial \Omega_{\mathrm{H}}}{\partial \sigma}=\frac{\partial \Omega_{\mathrm{H}}}{\partial \omega}=\frac{\partial \Omega_{\mathrm{H}}}{\partial \rho}= \frac{\partial \Omega_{\mathrm{H}}}{\partial a}=0.
\end{equation}
We also need to consider the $\beta$ equilibrium and the charge neutrality conditions,
\begin{align}
\mu_{e}=\mu_{\mu}=-\mu_{Q} ,\\
\frac{\partial \Omega_{\mathrm{H}}}{\partial \mu_{Q}}=n_{p}-n_{l}=0 \,,
\end{align}
where $\mu_Q$ is the charge chemical potential. 
We then have the pressure in hadronic matter as
\begin{equation}
P_{\mathrm{H}}=-\Omega_{\mathrm{H}}.
\end{equation}

The model parameters are determined by fitting to vacuum properties and nuclear saturation data. Tab.~\ref{input: mass} lists the hadron masses and pion decay constant used as inputs, while Tab.~\ref{saturation} summarizes the nuclear matter saturation properties at $n_0 = 0.16$ fm$^{-3}$.
\begin{table}[htbp]
\centering
	\caption{  {\small Physical inputs in vacuum in unit of MeV.  }  }\label{input: mass}
	\begin{tabular}{cccccccc}
		\hline\hline
		~$m_\pi$~&~ $f_\pi$ ~&~$m_\eta$ ~&~ $m_{a0}$ ~&~ $m_\omega$ ~&~ $m_\rho$ ~&~ $m_+$ ~&~ $m_-$\\
		\hline
		~140 ~&~ 92.4 ~&~ 550 ~&~ 980 ~&~ 783 ~&~ 776 ~&~ 939 ~&~ 1535\\
		\hline\hline
	\end{tabular}
\end{table}	
\begin{table}[htbp]
\centering
	\caption{  {\small Saturation properties used to determine the model parameters: the saturation density $n_0$, the binding energy $E_{\rm Bind}$, the incompressibility $K_0$, symmetry energy $S_0$. 
 }  }
	\begin{tabular}{cccc}\hline\hline
	~$n_0$ [fm$^{-3}$] ~& $E_{\rm Bind}$ [MeV] ~& $K_0$ [MeV] ~& $S_0$ [MeV] ~\\
	\hline
	0.16 & 16 & 240 & 31 \\
	\hline\hline
	\end{tabular}
	\label{saturation}
\end{table}	
Additionally, we constrain the vector meson mixing coefficient $\lambda_{\omega\rho}$ by requiring the slope parameter of the symmetry energy to match the empirical value $L = 57.7 \pm 19$ MeV~\cite{universe7060182}. In this work, we use the central value $L = 57.7$ MeV.

\begin{figure}[htbp]
\centering
\includegraphics[width=1\hsize]{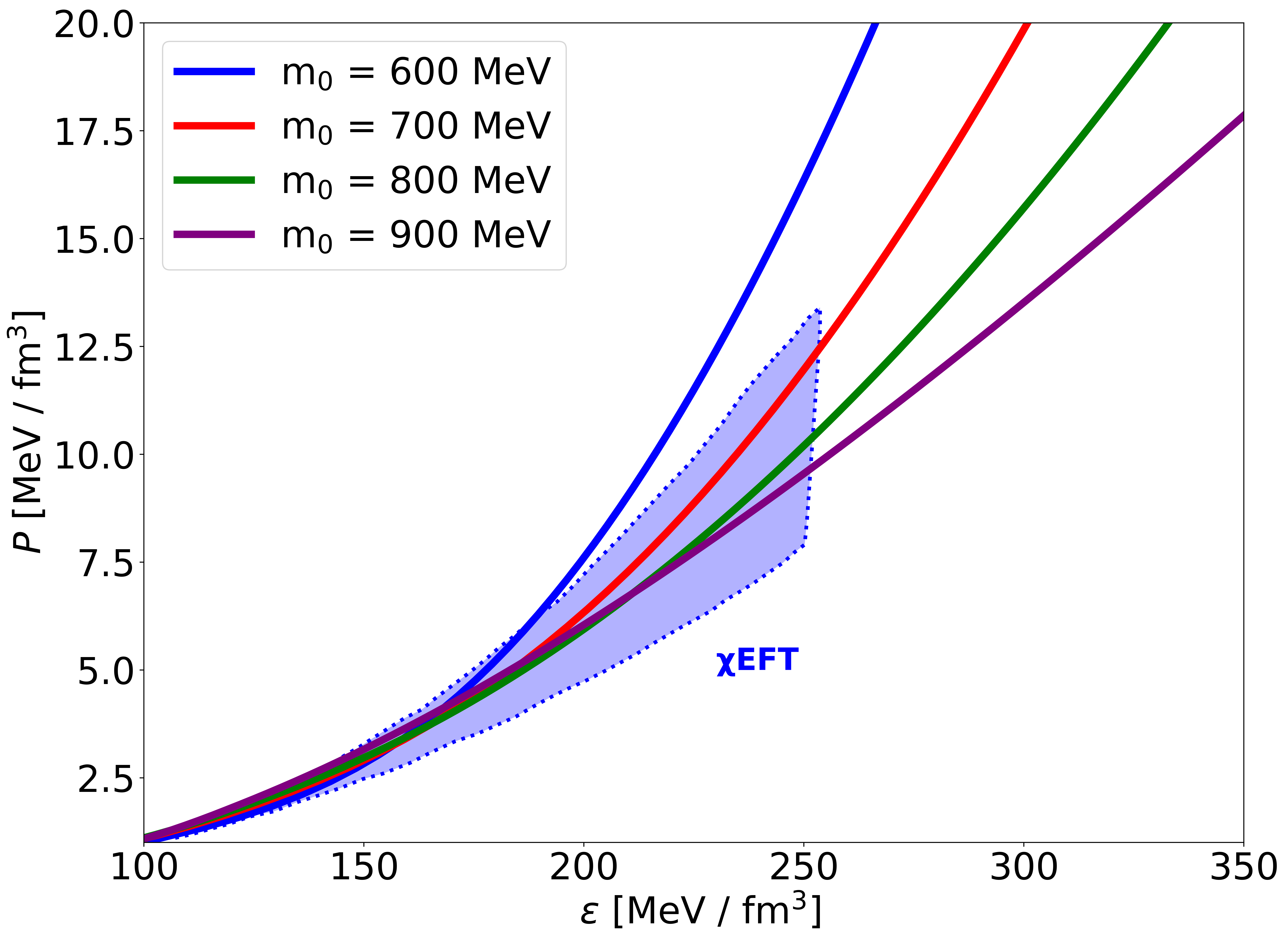}
\caption{EOS for different values of $m_0$ for $L=57.7$ MeV. The blue band denotes the 1$\sigma$ uncertainty in $\chi$EFT~\cite{Drischler:2017wtt,Drischler:2020fvz,Keller:2022crb}. }
\label{L_cons}
\end{figure}

\begin{table}[h]
\caption{Values of parameters $g_1, g_2, \bar{\mu}^2_{\sigma}, \bar{\mu}^2_{a}, \lambda_4, \gamma_4, \lambda_6, g_{\omega NN}, g_{\rho NN}$ and $\lambda_{\omega \rho}$ for $m_0=600$--$900$ MeV.}
\label{tab:parameters}
\centering
\begin{tabular}{l|cccc}
\hline\hline
~$m_0$ [MeV] ~&~~ 600 ~~&~~~ 700 ~~~& ~~~800 ~~~&~~~ 900 ~\\
\hline
$g_1$ & 8.48 & 7.81 & 6.99 & 5.96 \\
\hline
$g_2$ & 14.93 & 14.26 & 13.44 & 12.41 \\
\hline
$\bar{\mu}^2_{\sigma} / f^2_{\pi}$ & 22.43 & 19.38 & 12.06 & 1.64 \\
\hline
$\lambda_4$ & 40.40 & 35.51 & 23.21 & 4.56 \\
\hline
$\lambda_6 f^2_{\pi}$ & 15.75 & 13.90 & 8.93 & 0.69 \\
\hline
$g_{\omega NN}$ & 9.14 & 7.31 & 5.66 & 3.52 \\
\hline
$\bar{\mu}^2_{a} / f^2_{\pi}$ & $-10.77$ & $-13.82$ & $-21.15$ & $-31.56$ \\
\hline
$\gamma_4$ & 180.45 & 168.18 & 135.97 & 84.38 \\
\hline
$g_{\rho NN}$ & 14.84 & 13.08 & 11.63 & 10.24 \\
\hline
$\lambda_{\omega \rho}$ & 0.02 & 0.06 & 0.20 & 1.55 \\
\hline\hline
\end{tabular}
\end{table}
Tab.~\ref{tab:parameters} presents the resulting model parameters for different values of the chiral invariant mass $m_0$ ranging from 600 to 900 MeV.
We can then calculate the  EOS in the hadronic model  and the  corresponding EOS for PDM with fixing slope parameter $L=57.7$ MeV is shown in Fig.~\ref{L_cons}. 
From  this figure, 
we easily find that larger values of $m_0$ lead to softer EOSs.
This is understood as follows:
a greater $m_0$ leads to a weaker $\sigma$ coupling to nucleons, because a nucleon does not have to acquire its mass entirely from the $\sigma$ fields. The couplings to $\omega$ fields are also smaller because the repulsive contributions from $\omega$ fields must be balanced with attractive $\sigma$ contributions at the saturation density $n_0$. At densities larger than $n_0$, however, the $\sigma$ field reduces but the $\omega$ field increases, and these contributions are no longer balanced, affecting the stiffness of the EOS. We also compare our EOS with the theoretical constraint from N$^3$LO $\chi$EFT, shown as the blue band with $1\sigma$ uncertainty~\cite{Drischler:2017wtt,Drischler:2020fvz,Keller:2022crb}. For $m_0 = 600$ MeV, the EOS lies outside the $\chi$EFT band, whereas for $m_0 = 700$, $800$, and $900$ MeV, the EOSs fall within the predicted range. Increasing $m_0$ leads to relatively better agreement with $\chi$EFT predictions.

\subsection{QUARK MATTER EOS}
\label{NJL matter}

Following Refs.~\cite{Baym:2017whm,Baym:2019iky}, we use an NJL-type quark model to describe the quark matter in the high density region ($\geq 5 n_0$). 
The model includes three-flavor and U(1)$_A$ anomaly effects through the quark version of the KMT interaction. The coupling constants are chosen to be the Hatsuda-Kunihiro parameters 
which successfully reproduce the hadron phenomenology at low energy \cite{Baym:2017whm, Hatsuda:1994pi}: 
$G\Lambda^{2}=1.835, K\Lambda^{5}=9.29$ with $\Lambda=631.4\, \rm{MeV}$, see the definition below.
The couplings $g_{V}$ and $H$ characterize the strength of the vector repulsion and attractive diquark correlations whose range will be examined later 
when we discuss the NS constraints.

We can then write down the thermodynamic potential as
\be
\Omega_{\mathrm{CSC}}
& = &\, \Omega_{s}-\Omega_{s}\left[\sigma_{f}=\sigma_{f}^{0}, d_{j}=0, \mu_{q}=0\right] \\
& & {} +\Omega_{c}-\Omega_{c}\left[\sigma_{f}=\sigma_{f}^{0}, d_{j}=0\right],
\ee
where 
the subscript 0 is attached for the vacuum values, and
\begin{align}
&\Omega_{s}=-2 \sum_{i=1}^{18} \int^{\Lambda} \frac{d^{3} \mathbf{p}}{(2 \pi)^{3}} \frac{\epsilon_{i}}{2} \label{energy eigenvalue},\\
&\Omega_{c}=\sum_{i}\left(2 G \sigma_{i}^{2}+H d_{i}^{2}\right)-4 K \sigma_{u} \sigma_{d} \sigma_{s}-g_{V} n_{q}^{2},
\end{align}
with $\sigma_{f}$ being the chiral condensate for flavor-$f$ quark, $d_{j}$ denotes for diquark condensates, and $n_{q}$ denotes for the quark density. 
In Eq.~(\ref{energy eigenvalue}), $\epsilon_{i}$ are energy eigenvalues obtained from inverse propagator in Nambu-Gorkov bases
\begin{equation}
S^{-1}(k)=\left(\begin{array}{lc}
\gamma_{\mu} k^{\mu}-\hat{M}+\gamma^{0} \hat{\mu} & \gamma_{5} \sum_{i} \Delta_{i} R_{i} \\
-\gamma_{5} \sum_{i} \Delta_{i}^{*} R_{i} & \gamma_{\mu} k^{\mu}-\hat{M}-\gamma^{0} \hat{\mu}
\end{array}\right),
\end{equation}
where
\begin{equation}
\begin{aligned}
&M_{i} =m_{i}-4 G \sigma_{i}+K\left|\epsilon_{i j k}\right| \sigma_{j} \sigma_{k}, \\
&\Delta_{i} =-2 H d_{i} ,\\
&\hat{\mu} =\mu_{q}-2 g_{V} n_{q}+\mu_{3} \lambda_{3}+\mu_{8} \lambda_{8}+\mu_{Q} Q,\\
&(R_{1}, R_{2}, R_{3})=(\tau_{7}\lambda_{7}, \tau_{5}\lambda_{5}, \tau_{2}\lambda_{2}).
\end{aligned}
\end{equation}
$S^{-1}(k)$ is $72\times72$ matrix in terms of the color,
flavor, spin, and Nambu-Gorkov basis, which has 72 eigenvalues. $M_{u,d,s}$ are the constituent masses of $u, d, s$ quarks and $\Delta_{1,2,3}$ are the gap energies. 
The $\mu_{3,8}$ are the color chemical potentials which will be tuned to achieve the color neutrality. 
The total thermodynamic potential including the effect of leptons is 
\begin{equation}
\Omega_{\mathrm{Q}}=\Omega_{\mathrm{CSC}}+\sum_{l=e, \mu} \Omega_{l}.
\end{equation}
The mean fields are determined from the gap equations,
\begin{equation}
0=\frac{\partial \Omega_{\mathrm{Q}}}{\partial \sigma_{i}}=\frac{\partial \Omega_{\mathrm{Q}}}{\partial d_{i}}.
\end{equation}
From the conditions for electromagnetic charge neutrality and color charge neutrality, we have
\begin{equation}
n_{j}=-\frac{\partial \Omega_{\mathrm{Q}}}{\partial \mu_{j}}=0,
\end{equation}
where $j = 3,8, Q$. 
The baryon number density $n_{B}$ is determined as
\begin{equation}
n_{q}=-\frac{\partial \Omega_{\mathrm{Q}}}{\partial \mu_{q}},
\end{equation}
where $\mu_{q}$ is $1/3$ of the baryon number chemical potential. After determined all the values, we obtain the pressure as
\begin{equation}
P_{\mathrm{Q}}=-\Omega_{\mathrm{Q}}.
\end{equation}
The remaining two parameters $H$ and $g_V$ will be constrained in the following sections.

\section{STUDY OF PROPERTIES OF NS}
\label{sec-NS}

Now, following Ref.~\cite{Baym:2017whm, Baym:2019iky, Minamikawa:2020jfj,Kojo:2021wax, Gao:2022klm, Minamikawa:2023eky} 
we are ready to construct a unified EOS by connecting the EOS obtained in the PDM introduced in Sec.~\ref{sec:PDM matter} 
and the EOS of NJL-type quark model given in Sec.~\ref{NJL matter}, and solve the TOV equation~\cite{Tolman:1939jz,Oppenheimer:1939ne} 
to obtain the NS mass-radius  ($M$-$R$)  relation. 
As for the interplay between nuclear and quark matter EOS, see, e.g., Ref.~\cite{Kojo:2020krb} for a quick review that classifies types of the interplay.


\subsection{Construction of unified EOS}
\begin{figure}[htbp]
\centering
\includegraphics[width=1\hsize]{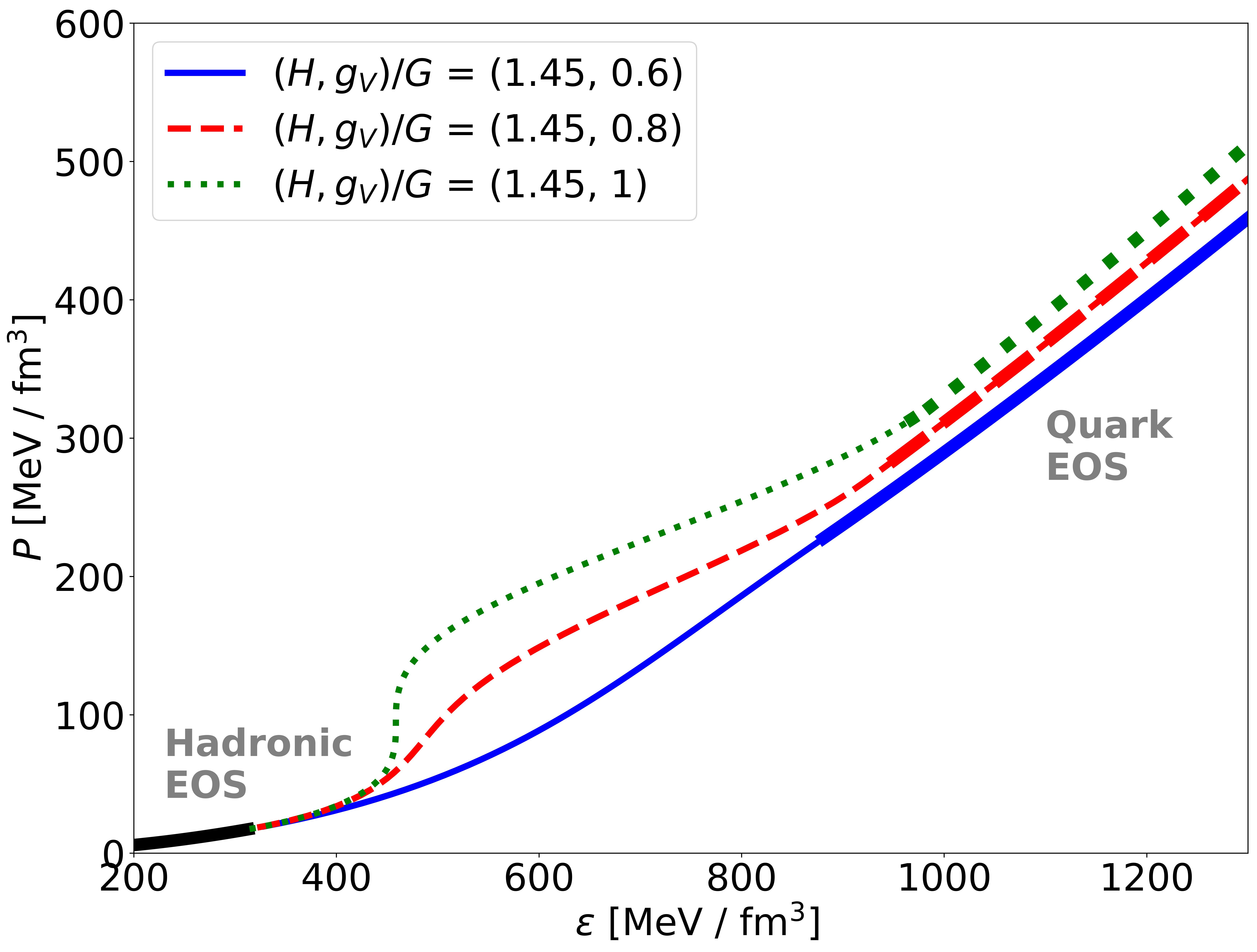}
\includegraphics[width=1\hsize]{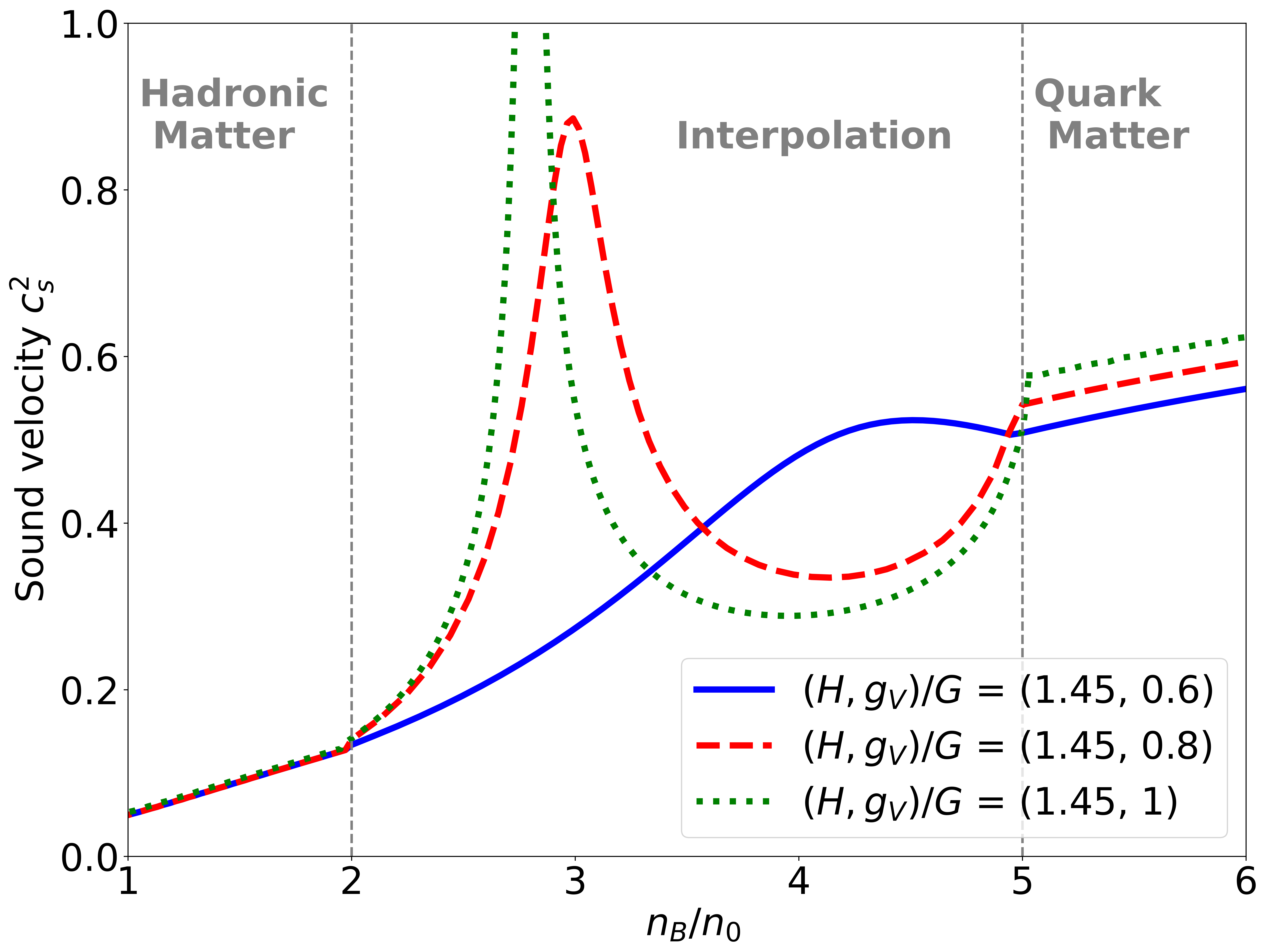}
\caption{
Upper panel: Unified EOS with $m_0 = 800$~MeV (black thick curve) and the choices of the NJL-type quark model for $(H, g_V)/G$ as $(1.45, 0.6)$ (blue thick curve); $(1.45, 0.8)$ (red thick curve), and $(1.45, 1.0)$ (green thick curve). The interpolated region is shown by thin curves. 
Lower panel: Corresponding sound velocity for the EOS shown in the upper panel.
}
\label{fig:interpolate}
\end{figure}

%
\begin{table}[tbh]
\begin{center}
\begin{tabular}{c|c|c|c}
\hline
\hline
$0\leq \bar{n}<0.5  $ & $0.5\leq \bar{n}\leq 2$ & $2<\bar{n}<5$ & $\bar{n}\geq 5$\\
\hline
\rm{Crust} & \rm{PDM} & \rm{Interpolation} & \rm{NJL}\\
\hline
\hline
\end{tabular}
\end{center}
\caption{Segmentation and composition of the unified EOS with normalized density $\bar{n}=n_B/n_0$.}
\label{UniEOS}
\end{table}
In our unified equations of state as in Tab.~\ref{UniEOS},
we use the BPS (Baym-Pethick-Sutherland) EOS~\cite{Baym:1971pw} as a crust EOS for $n_B \lesssim 0.5n_0$. 
From $n_B \simeq 0.5n_0$ to $2n_0$ we use our PDM model to describe the nuclear matter.  
We limit the use of our PDM up to $2n_{0}$ so that baryons other than ground state nucleons, such as the negative parity nucleons or hyperons, do not show up. Beyond $2n_0$ nuclear regime, we assume a crossover from the nuclear matter to quark matter, 
and use a smooth interpolation to construct the unified EOS. 
We expand the pressure as a fifth order polynomial of $\mu_{B}$ as
\begin{equation}
P_{\mathrm{I}}\left(\mu_{B}\right)=\sum_{i=0}^{5} C_{i} \mu_{B}^{i},
\end{equation}
where $C_{i}$  ($i=0,\cdots, 5$) are parameters  to be determined from boundary conditions  given by 
\begin{equation}
\begin{aligned}
&\left.\frac{\mathrm{d}^{n} P_{\mathrm{I}}}{\left(\mathrm{d} \mu_{B}\right)^{n}}\right|_{\mu_{B L}}=\left.\frac{\mathrm{d}^{n} P_{\mathrm{H}}}{\left(\mathrm{d} \mu_{B}\right)^{n}}\right|_{\mu_{B L}}, \\
&\left.\frac{\mathrm{d}^{n} P_{\mathrm{I}}}{\left(\mathrm{d} \mu_{B}\right)^{n}}\right|_{\mu_{B U}}=\left.\frac{\mathrm{d}^{n} P_{\mathrm{Q}}}{\left(\mathrm{d} \mu_{B}\right)^{n}}\right|_{\mu_{B U}}, \quad(n=0,1,2),
\end{aligned}
\end{equation}
with $\mu_{BL}$  being 
the chemical potential corresponding to $n_{L}=2n_{0}$ and $\mu_{BU}$ to $n_{U}=5n_{0}$. We demand the matching up to the second order derivatives of pressure at each boundary.
The resultant interpolated EOS must satisfy the thermodynamic stability condition,
\begin{align}
\chi_B = \frac{\, \partial^2 P \,}{\, (\partial \mu_B )^2 \,} \ge 0 \,,
\end{align}
and the causality condition,
\begin{equation}
c_{s}^{2}=\frac{\, \mathrm{d} P \,}{\mathrm{d} \varepsilon} 
= \frac{n_{B}}{\mu_{B} \chi_{B}} \le 1 \,,
\end{equation}
which means that the sound velocity is smaller than the speed of light. Figure~\ref{fig:interpolate} shows the interpolated EOS. 
In the upper panel, hadronic matter is described by the PDM with $m_0 = 800$~MeV (black thick curve) and smoothly interpolated with the quark matter EOS obtained from the NJL-type quark model for $(H, g_V)/G$ values of (1.45, 0.6) --- blue thick curve; (1.45, 0.8) --- red thick curve; and (1.45, 1.0) --- green thick curve. 
The thin curves indicate the interpolation region. 
As the vector repulsive interaction parameter $g_V$ increases, the quark EOS becomes stiffer, and this stiffness is also reflected in the interpolated EOS. 
The lower panel shows the corresponding sound velocity. 
The maximum value of $c_s^2$ increases with $g_V$, and for $(H, g_V)/G = (1.45, 1.0)$ it exceeds 1, indicating a violation of causality; therefore, such parameter choices should be excluded.

These conditions restrict the range of quark model parameters $(g_V, H)$ for a given nuclear EOS and a choice of $(n_L, n_U)$. We exclude interpolated EOSs which do not satisfy the above-mentioned constraints. We note that the interpolation between the hadronic (PDM) and quark (NJL) models is performed over a relatively wide density range (2-5 times normal nuclear density) because there is still significant uncertainty regarding the exact density at which the transition from hadronic to quark matter occurs in neutron stars. 

\subsection{Mass-Radius relation}

In this section, we calculate the mass-radius ($M$-$R$) relations of neutron stars using the unified EOSs constructed in the previous section, examining how different values of the chiral invariant mass $m_0$ affect NS properties.
\begin{figure*}[t]\centering
\begin{subfigure}{0.47\hsize}\centering
	\includegraphics[width=0.95\hsize]{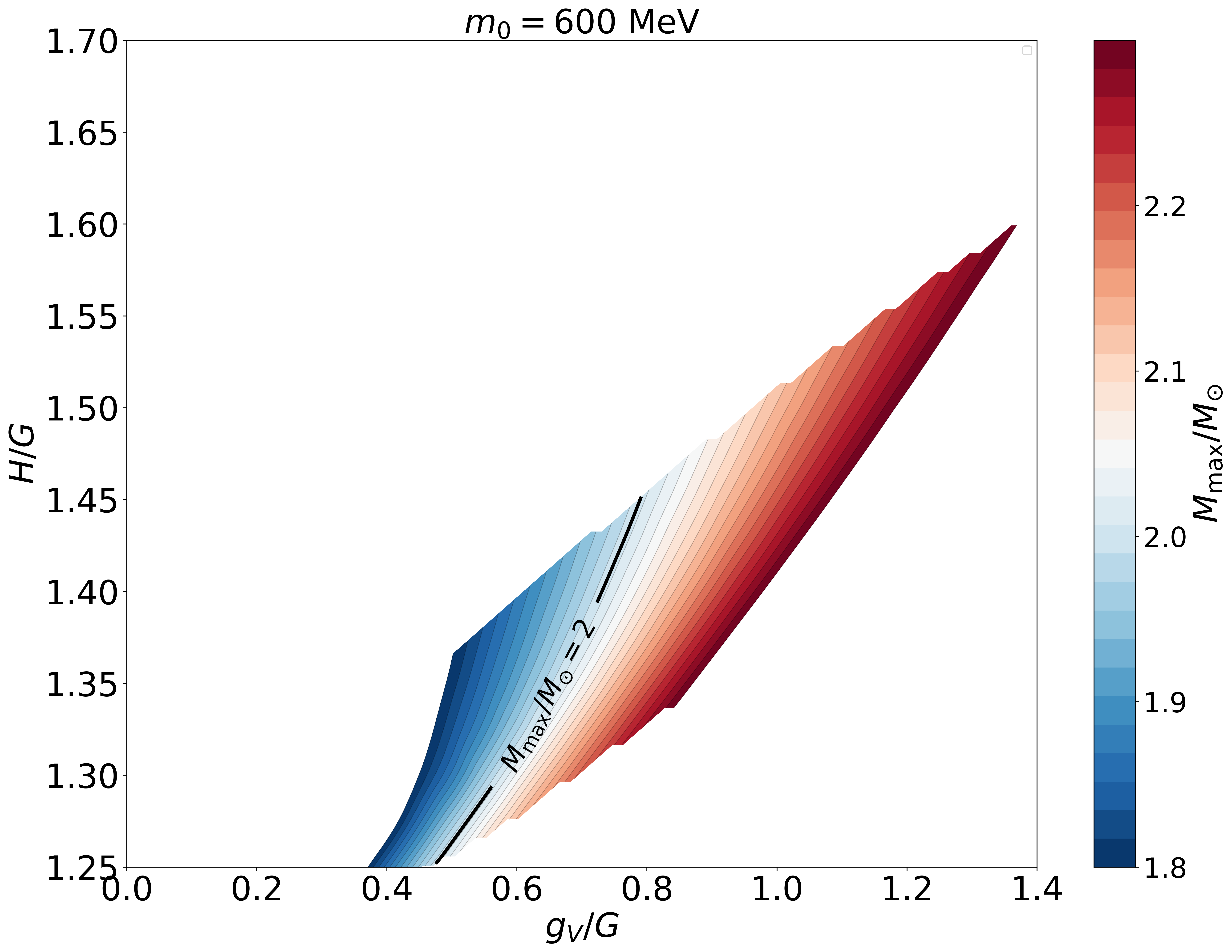}
\end{subfigure}
\begin{subfigure}{0.47\hsize}\centering
	\includegraphics[width=0.95\hsize]{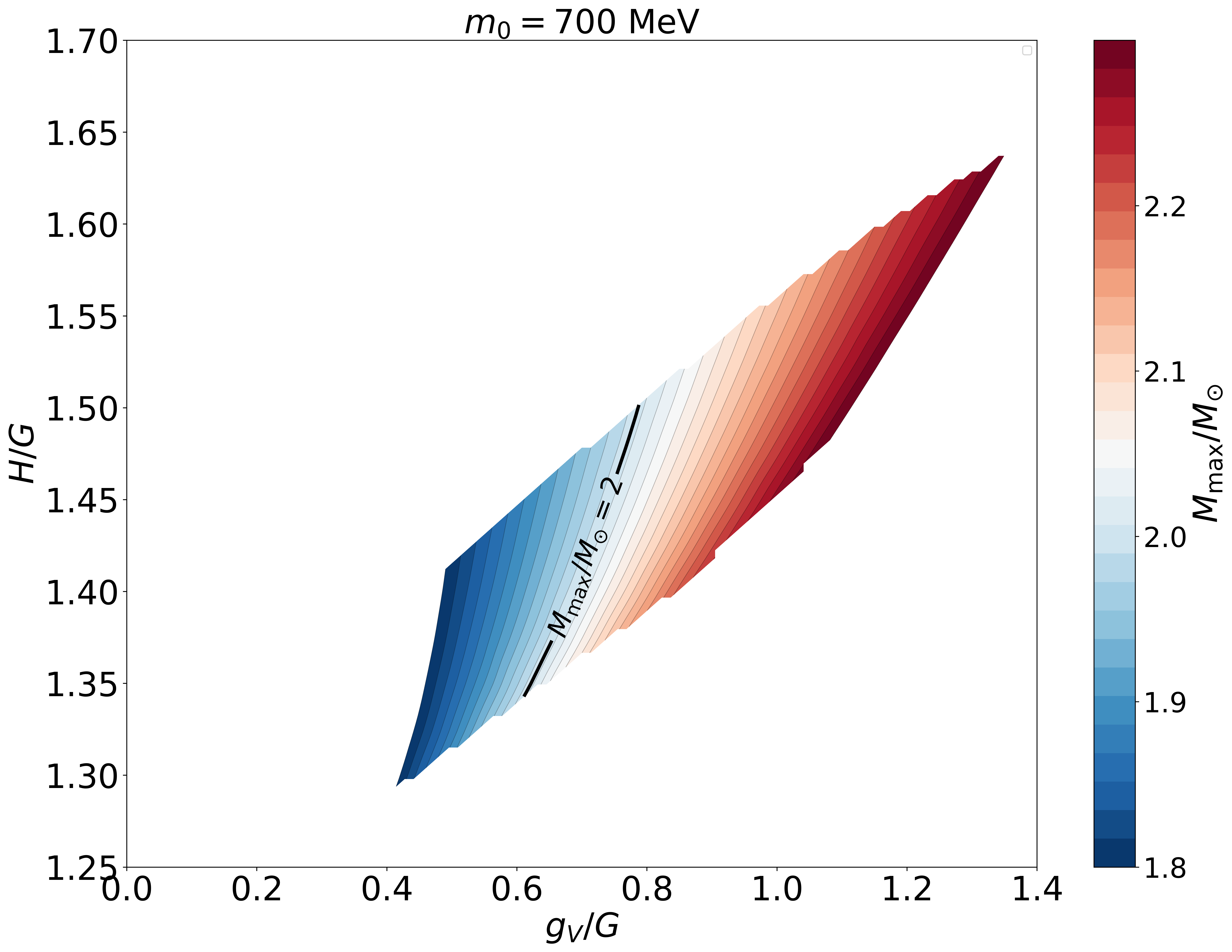}
\end{subfigure}
\begin{subfigure}{0.47\hsize}\centering
	\includegraphics[width=0.95\hsize]{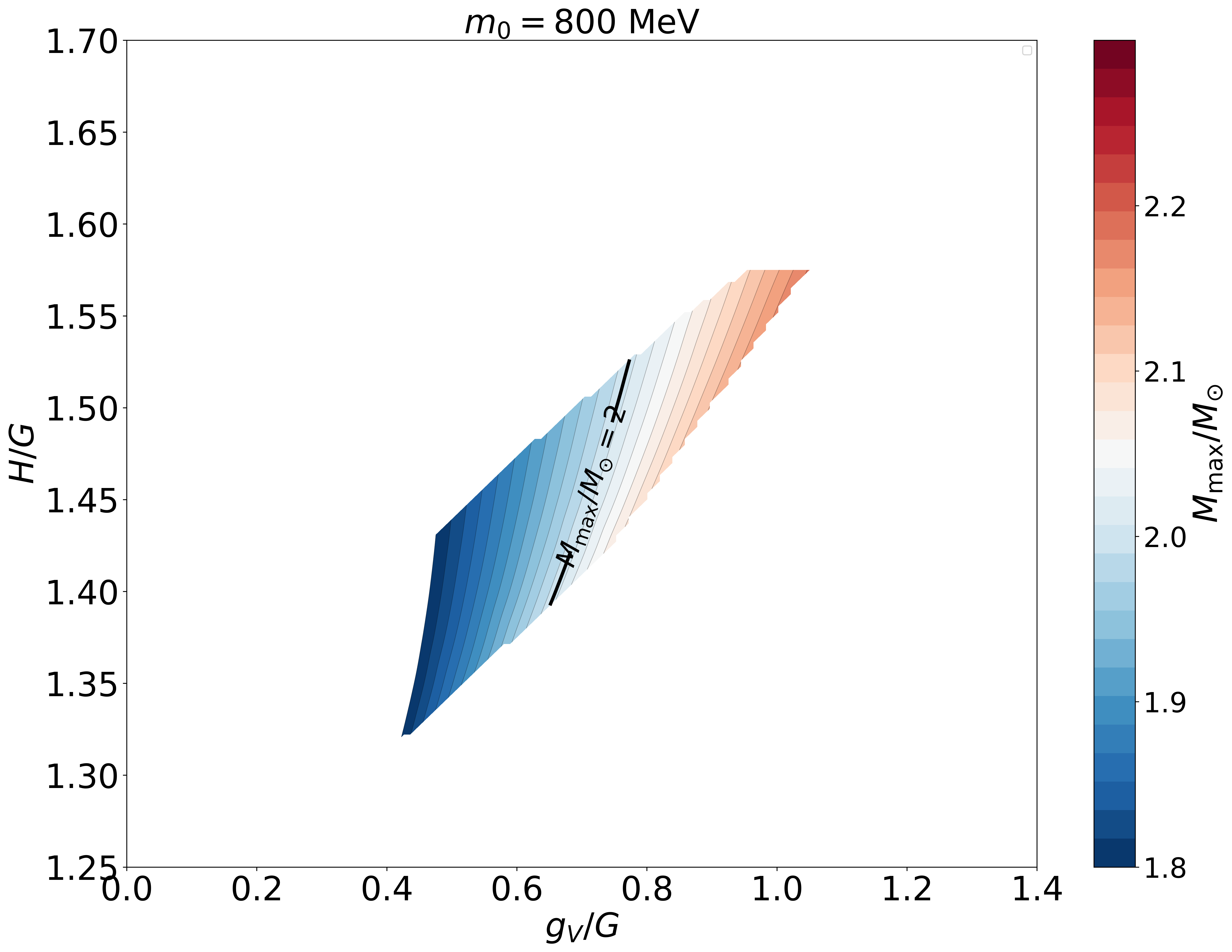}
\end{subfigure}
\begin{subfigure}{0.47\hsize}\centering
	\includegraphics[width=0.95\hsize]{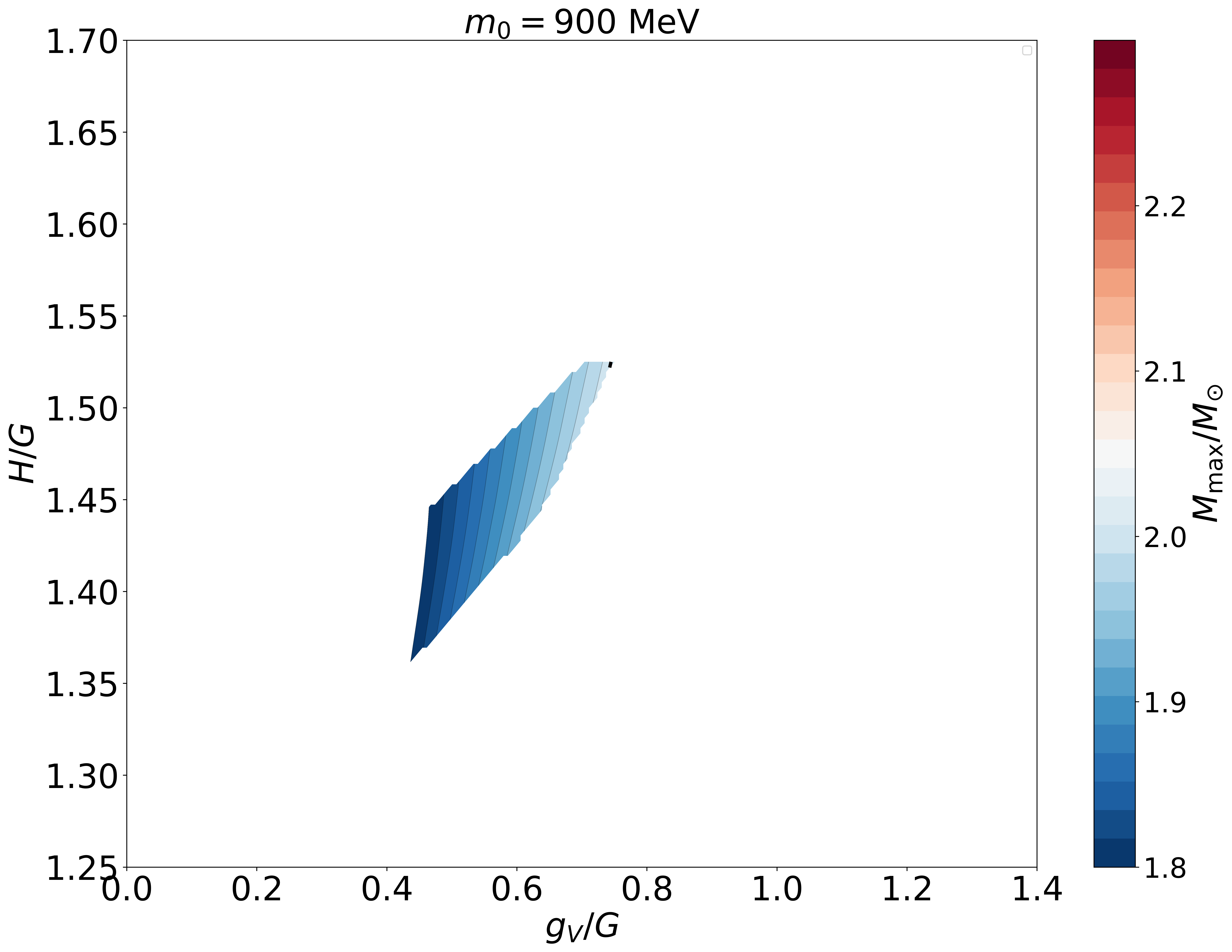}
\end{subfigure}
\caption[]{
Allowed combination of ($H, g_V$) values for $m_0= 600, 700, 800, 900$ MeV when $L=57.7$ MeV. The uncolored region indicates that the combination of ($H, g_V$) is excluded by the causality constraints. The colored region indicates that the combination is allowed. The color shows the maximum mass of NS obtained from the corresponding parameters, as indicated by a vertical bar at the right side of each figure.}
\label{H_gv}
\end{figure*}
For each hadronic EOS characterized by a fixed value of $m_0$, we construct unified EOSs by connecting to quark matter described by different combinations of the diquark coupling $H$ and vector repulsion strength $g_V$. The viable parameter space for $(H, g_V)$ is first constrained by two fundamental requirements: causality and the existence of massive neutron stars.

Figure~\ref{H_gv} illustrates the allowed parameter space for different values of $m_0 = 600, 700, 800, 900$ MeV. The uncolored regions indicate parameter combinations where the interpolated EOS violates causality (i.e., the sound speed exceeds the speed of light) and must be excluded. In the colored regions that satisfy causality, the color coding represents the maximum NS mass achievable for each $(H, g_V)$ combination. The black curves mark parameter combinations yielding a maximum mass of exactly $2M_{\odot}$, the observational lower bound from massive pulsars. Parameter combinations below these curves are ruled out by observations. The parameter dependence shows a clear trend: increasing $g_V$ (stronger vector repulsion) or decreasing $H$ (weaker diquark pairing) leads to higher maximum masses.

\begin{figure}[htbp]
\centering
\includegraphics[width=1\hsize]{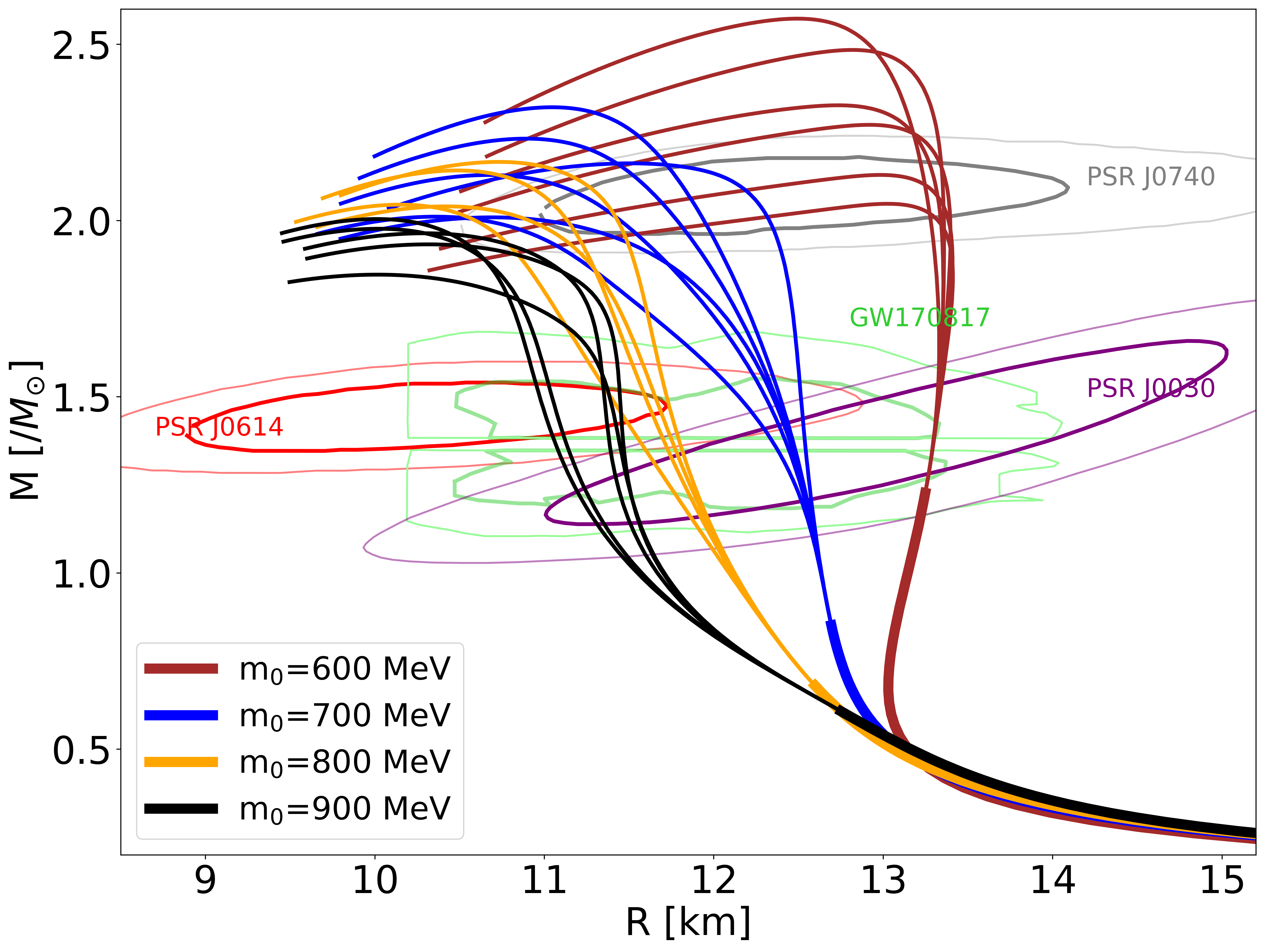}
\caption{
$M$–$R$ relation for unified EOS with several choices of $m_0$, interpolated with different quark matter EOS. The thick curves indicate the density region where the hadronic EOS is applied (up to $2n_0$) for $m_0 = 600$ MeV (brown curve), $700$ MeV (blue curve), $800$ MeV (orange curve), and $900$ MeV (black curve). Observational constraints are shown as contours from PSR J0740+6620 (grey), PSR J0030+0451 (purple), GW170817 (green), and the latest measurement of PSR J0614-3329 (red).
}
\label{fig:mr}
\end{figure}

Having constrained the parameter space through maximum mass requirements, we now examine how NS radii provide additional constraints. Figure~\ref{fig:mr} presents the resulting $M$-$R$ relations for unified EOSs with different values of $m_0$, each connected to various quark matter EOSs. The thick portions of the curves indicate where the hadronic PDM is applied (up to $2n_0$), shown for $m_0 = 600$ MeV (brown), 700 MeV (blue), 800 MeV (orange), and 900 MeV (black). The observational constraints are displayed as contours from PSR J0740+6620 (grey), PSR J0030+0451 (purple), GW170817 (green), and the recent PSR J0614-3329 (red).

As expected from our EOS analysis, larger values of $m_0$ lead to softer EOSs, resulting in smaller NS radii. The comparison with observations reveals distinct patterns for each $m_0$ value. For $m_0 = 600$ MeV, while the unified EOS remains consistent with GW170817, PSR J0030+0451, and PSR J0740+6620, it lies outside the PSR J0614-3329 constraint, producing radii that are too large  to satisfy all the observations. For $m_0 = 700$ MeV, the corresponding $M$-$R$ curves fall within the 1$\sigma$ regions of GW170817, PSR J0030+0451, and PSR J0740+6620, though only achieving consistency with PSR J0614-3329 at the 2$\sigma$ level. For $m_0 = 800$ MeV, it satisfies all observational constraints within their 1$\sigma$ uncertainties. However, further increasing to $m_0 = 900$ MeV creates tension in the opposite direction: although this value produces sufficiently small radii to satisfy GW170817, PSR J0030+0451, and PSR J0614-3329, it generates radii too small to accommodate the PSR J0740+6620 constraint, demonstrating that excessively large chiral invariant masses are also disfavored.

These observational constraints significantly narrow the allowed range of the chiral invariant mass. Prior to the PSR J0614-3329 measurement, previous analyses~\cite{PhysRevC.108.055206, Kong:2025dwl} had constrained the chiral invariant mass to
\begin{align}
580~\text{MeV} \lesssim m_0 \lesssim 860~\text{MeV}.
\end{align}
The inclusion of PSR J0614-3329, with its unprecedentedly small radius, substantially tightens this constraint to
\begin{align}
800~\text{MeV} \lesssim m_0 \lesssim 860~\text{MeV},
\end{align}
raising the lower bound by approximately 220 MeV. This dramatic refinement underscores the importance of PSR J0614-3329 in constraining fundamental properties of dense matter and highlights that the chiral invariant mass must constitute a significant fraction ($\gtrsim 85\%$) of the nucleon mass.

\section{SUMMARY AND DISCUSSIONS}
\label{summary}

In this work, we have investigated the implications of the recent NICER observation of PSR J0614-3329 for the chiral invariant mass $m_0$ within the parity doublet model framework. This pulsar, with its unprecedentedly small radius of $R = 10.29^{+1.01}_{-0.86}$ km at mass $M = 1.44^{+0.06}_{-0.07} M_\odot$, provides the most stringent constraint on neutron star radii to date and offers a unique opportunity to probe fundamental properties of dense QCD matter.

We constructed unified equations of state by employing the parity doublet model with isovector scalar meson $a_0(980)$ contributions for the hadronic phase up to $2n_0$, smoothly connected to an NJL-type quark model at higher densities through a crossover transition. The PDM naturally incorporates both the density-dependent chiral variant mass arising from spontaneous chiral symmetry breaking and the chiral invariant mass $m_0$ that persists even in the chirally restored phase. By systematically varying $m_0$ from 600 to 900 MeV and exploring different quark matter parameters $(H, g_V)$, we examined which parameter combinations satisfy the complete set of current astrophysical constraints.

Our analysis reveals that the inclusion of PSR J0614-3329 dramatically refines the allowed range of the chiral invariant mass. While previous studies without this constraint had established $580~\text{MeV} \lesssim m_0 \lesssim 860~\text{MeV}$, the requirement to simultaneously satisfy PSR J0614-3329's small radius and the existence of $2M_\odot$ neutron stars narrows this range to $800~\text{MeV} \lesssim m_0 \lesssim 860~\text{MeV}$. This represents an increase of approximately 220 MeV in the lower bound, demonstrating the powerful constraining capability of this single observation.

The refined constraint has profound implications for our understanding of the nucleon mass origin. Our results indicate that the chiral invariant mass must constitute at least 85\% of the nucleon mass, challenging the traditional picture where the nucleon mass emerges predominantly from spontaneous chiral symmetry breaking. This finding aligns with recent lattice QCD simulations, QCD sum rule and skyrmion crystal calculations that suggest a significant fraction of the nucleon mass persists even in the chiral restored phase. The large chiral invariant mass required by astrophysical observations implies that gluon condensation and other non-chiral mechanisms play a more dominant role in generating the nucleon mass than previously anticipated.

Finally, we note that the constraints on the chiral invariant mass are not so sensitive to the choice of the interpolation boundary $(n_L, n_U)$. Explicitly, we examined two variations. First, when $n_L$ is decreased to $1.5n_0$, allowing the hadronic description to extend to lower densities, the constrained region expands to approximately $790~\text{MeV} \lesssim m_0 \lesssim m_N$. Oppositely, when $n_L$ is increased to $2.5n_0$, the allowed $(H, g_V)$ parameter space shrinks, yielding a tighter constraint of $800~\text{MeV} \lesssim m_0 \lesssim 815~\text{MeV}$. However, for such large values of $n_L$, the reliability of the hadronic description becomes questionable, particularly given the uncertainties in hyperon appearance at these densities.
These results demonstrate that while the upper bound of $m_0$ shows some sensitivity to the interpolation boundary, the lower bound remains relatively stable around $790$--$800~\text{MeV}$. The observation of PSR J0614-3329 provides a significant improvement over our previous constraints on the chiral invariant mass.

Looking forward, future observations of compact neutron stars with small radii, combined with continued detection of massive pulsars, will further constrain the chiral invariant mass and deepen our understanding of QCD in the non-perturbative regime. The tight correlation between astrophysical observations and fundamental QCD parameters demonstrated in this work underscores the unique role of neutron stars as laboratories for studying the strong interaction under extreme conditions. As multi-messenger astronomy continues to advance with improved sensitivity from NICER, gravitational wave detectors, and other observational facilities, we anticipate even more precise determinations of the fundamental parameters governing dense matter, ultimately revealing the true nature of nucleon mass generation and chiral symmetry restoration in QCD.

\section*{Acknowledgement}
This work of Y. K. Kong was supported by JST SPRING, Grant Number
JPMJSP2125. Y. L. M. was supported in part by the National Science Foundation of China (NSFC) under Grant No. 12347103, the National Key R\&D Program of China under Grant No. 2021YFC2202900 and Gusu Talent In novation Program under Grant No. ZXL2024363. Y. K. Kong would like to take this opportunity to thank the
“THERS Make New Standards Program for the Next Generation Researchers.”

\bibliography{ref_3fPDM_2022.bib}

\end{document}